\documentclass[prl,aps,twocolumn,floats,superscriptaddress,showpacs]{revtex4}

\usepackage{graphics}
\usepackage{amsmath}
\usepackage{dcolumn}
\usepackage{citesort}
\begin{document}

\title{{\rm\small\hfill (Phys. Rev. Lett., accepted)}\\
Towards an exact treatment of exchange and correlation in materials:\\
Application to the ``CO adsorption puzzle'' and other systems}

\author{Qing-Miao Hu}

\altaffiliation[Permanent address: ]{Institute of Metal Research - CAS, Shenyang 110016, P.R. China}

\affiliation{Fritz-Haber-Institut der Max-Planck-Gesellschaft,
Faradayweg 4-6, D-14195 Berlin, Germany}

\author{Karsten Reuter}

\affiliation{Fritz-Haber-Institut der Max-Planck-Gesellschaft,
Faradayweg 4-6, D-14195 Berlin, Germany}

\author{Matthias Scheffler}

\affiliation{Fritz-Haber-Institut der Max-Planck-Gesellschaft,
Faradayweg 4-6, D-14195 Berlin, Germany}

\affiliation{Departments of Chemistry \& Biochemistry and Materials,
UCSB, Santa Barbara, CA 93106, USA}

\begin{abstract}
It is shown that the errors of present-day exchange-correlation (xc)
functionals are rather short ranged. For extended systems the
\textit{correction} can therefore be evaluated by analyzing properly
chosen clusters and employing highest-quality quantum chemistry methods.
The xc \textit{correction} rapidly approaches a universal dependence
with cluster size. The method
is applicable to bulk systems as well as to defects in the bulk and at
surfaces. It is demonstrated here for CO adsorption at transition-metal
surfaces, where present-day xc functionals dramatically
fail to predict the
correct adsorption site, and for the crystal bulk cohesive energy.
\end{abstract}


\maketitle

Electronic structure theory is the base for a multiscale modeling of
materials properties and functions (see e.g. Ref.
\cite{Reuter-SY-2005}). Obviously, if the needed accuracy is lacking at
this base, there is little hope that accurate predictions can be made at
any level of modeling that follows. For polyatomic systems,
density-functional theory (DFT) with present-day exchange-correlation
(xc) functionals has proven to be an excellent technique for
calculations at this electronic-structure base. However, it is not as
good for certain types of binding interactions. Accurate treatments of
strong electronic correlations, van der Waals interactions, and
molecular dynamics for electronically excited states
represent unsolved challenges. Besides numerical approximations
(e.g. the basis set, possible use of the pseudopotential approximation,
etc.), that good theoretical work is typically scrutinizing, a
satisfying test of the quality of the xc functional was
not possible for bigger systems, so far. Sometimes the results obtained
with different functionals have been compared, and when they agreed this
was taken as indicator for reliability. Though this was the best
possible approach, it is neither safe nor justified.
Exchange-correlation functionals are typically built on the homogeneous
electron gas [the local-density approximation (LDA) \cite{perdew92}],
adding corrections while ensuring consistency with known sum
rules [e.g. the generalized gradient approximation (GGA)
\cite{perdew96}], or they are constructed to reproduce certain data of
some small molecules (e.g. the B3LYP functional \cite{b3lyp}). There is
no systematic expansion in terms of successively decreasing errors, and, 
there is no proof that, e.g. the GGA will always work more trustfully 
than the LDA. On the other hand, for wavefunction methods, several
promising concepts exist for better xc treatments also for 
extended systems (see Ref. \cite{Fulde, Paulus} and references therein). 
However, these are not yet efficient, in particular for metals.

It is often argued that the xc error is largely canceled when
total-energy \textit{differences} are studied, and that the xc
approximation affects the geometry only by little. However, this is
generally not correct. A key example for the xc problem is the
low-coverage adsorption of CO at the (111) surface of Pt or Cu, where
the LDA as well as the GGA predict that the molecule adsorbs in the
threefold-coordinated hollow site. Experiments, on the other hand, show
undoubtedly that the adsorption site is in the one-fold coordinated top
site. Obviously the conclusion based on DFT-LDA/GGA is even
qualitatively incorrect; and when comparing the calculated energies of
the two sites in question, the error in the energy \textit{difference}
is indeed significant: In the LDA it must be larger than 0.4 eV.

The comprehensive study of CO at Pt(111) by Feibelman {\em et al.}
\cite{Feibelman-puzzle-2001} set the ball rolling, showing that when
properly realizing all technical aspects of the calculations (e.g. basis
set, supercell, cluster geometry) the LDA and GGA put the molecule at
the wrong adsorption site. For several other close packed transition
metal surfaces, including Cu(111), the situation is analogous (e.g.
\cite{Gajdos-2004}). Shortly after the paper by Feibelman \textit{et
al.} \cite{Feibelman-puzzle-2001} it was realized that the wrong site
preference of CO may be related to the fact that DFT-LDA and GGA
inaccurately describe the CO-molecule's chemical bond. This was then
expressed in terms of the HOMO-LUMO splitting (the 5$\sigma$ and
2$\pi^*$ energy levels) or correspondingly the singlet-triplet
excitation.\cite{Grinberg-2002,Olsen-2003,Gil-2003,Mason-2004,Gajdos-2005}
An upshift of the one electron 2$\pi^*$ level makes a partial charge
transfer from the 5$\sigma$ to the 2$\pi^*$ orbital more difficult and
therefore stabilizes the CO bond. Indeed, such (semi-empirical) upshift
brought the CO molecule to the experimentally known site. Thus, the 
problem appears to be understood (to some degree), but altogether the
situation is unsatisfactory. A first-principles theory should provide
a reliable answer, and an add-on, that is triggered by a disagreement
with experiment, questions the usefulness of the whole 
self-consistent procedure. We also note that the full correction of
the LDA/GGA functional will not just shift the CO $2\pi^*$ orbital
to higher energies; it will also modify the substrate 
\textit{d}-states and thereby the substrate-adsorbate bonding in
various terms. Below we will take the ``CO adsorption puzzle'' as 
our main example for a systematic, non-empirical approach to correct
xc errors. And we will also discuss a fundamentally different case,
namely the bulk cohesive energy.

For small systems (say up to 20-50 atoms) high-level quantum chemistry
methods or the quantum Monte Carlo approach can be employed to obtain
accurate results for the exchange-correlation energy. We will show that
calculations on such small clusters are sufficient to evaluate the
DFT-LDA \textit{error} of extended systems. We will concentrate on
CO/Cu(111), CO/Ag(111), and Cu bulk as for these systems relativistic
effects are small. We use accurate full-potential augmented
plane wave (LAPW/APW+lo) DFT calculations \cite{wien2k} with LDA
\cite{perdew92} and GGA \cite{perdew96} functionals to calculate the CO
adsorption energies with a numerical uncertainty of $\pm 0.02$\,eV
\cite{basis}. Supercells with relaxed, symmetric five layer slabs are
employed to model the low-coverage adsorption at 1/9 monolayer on the
extended surface. In good agreement with previous 
studies \cite{Gajdos-2004,Mason-2004,Gajdos-2005}, we obtain the fcc
hollow site to be more stable than the correct top site (by 0.33\,eV
within the LDA, by 0.11\,eV within the GGA).

Our approach works as follows (we are using here DFT-LDA as the starting
point, but, of course, one could as well start with other functionals,
e.g. DFT-GGA):

1. Do supercell calculations for the extended system using DFT-LDA.

2. Do cluster calculations with the same functional and same geometry as
in step 1. It may be convenient to saturate the cluster surface by
hydrogen adatoms, but there is in principle no need to do so.
	
3. Do corresponding calculations for exactly the same cluster as in step
2, but using an improved xc treatment. This may employ the B3LYP
functional, or Hartree-Fock (HF) plus M{\o}ller-Plesset perturbation
theory (MP2), a coupled-cluster, or a quantum Monte Carlo calculation.

\begin{figure}
\scalebox{0.77}{\includegraphics{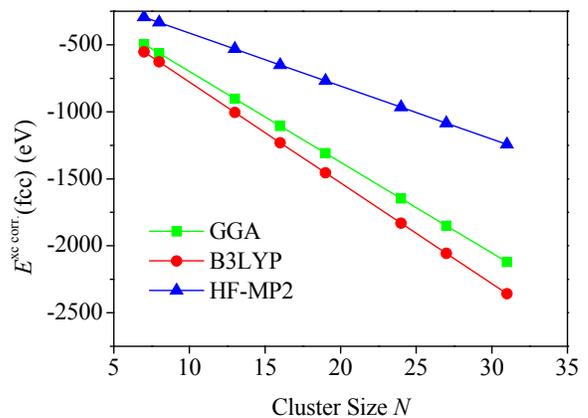}}
\caption{
Total energy correction $E^{\rm xc\,corr.}$, see eq. (\ref{xc-corr.}),
with respect to the LDA as function of cluster size and for xc = GGA,
B3LYP, and HF-MP2 for the adsorption of CO at Cu(111) in the fcc hollow
site.}
\label{fig-xc-corr.}
\end{figure}

The difference of the results of steps 3 and 2, i.e.
\begin{equation}
\label{xc-corr.}
E^{\rm xc\,corr.} = E^{\rm cluster}({\rm xc-better}) - E^{\rm
cluster}({\rm LDA})\quad,
\end{equation}
then is the xc correction. Why should this cluster quantity, $E^{\rm
xc\,corr.}$ be of much relevance for the extended system? Figure 
\ref{fig-xc-corr.} shows how $E^{\rm xc\,corr.}$ changes with cluster size
using the CO adsorption energy in the fcc hollow site at Cu(111) as
example. These calculations \cite{gaussian03,basis2} were performed at
the DFT-LDA, GGA, and B3LYP levels, as well as with HF-MP2.
The correction of the LDA result is dramatic: It is of the order of several 
hundreds eV, and the different xc treatments all give noticeably
different results. In this respect we note that none of the employed
methods (LDA, GGA, B3LYP, HF-MP2) fulfills the variational principle for
the ground state of the true many-electron Hamiltonian, i.e. they all
could, in principle, give results that are even below the true many-body
ground state energy. For a free CO molecule, for example, the total
energies are $-3\,059.01$\,eV (LDA), $-3\,079.83$\,eV (GGA),
$-3\,083.30$\,eV (B3LYP), $-3\,075.89$\,eV (HF-MP2), and the
experimental value is $-3\,084.72$\,eV.\cite{exp.data} Thus, the order
is LDA $>$ HF-MP2 $>$ GGA $>$ B3LYP $>$ experiment, were the differences
to the experimental values are between $-25.7$\,eV (LDA) and $-1.4$\,eV
(B3LYP). The trend seen in Fig. \ref{fig-xc-corr.} is thus the same as
that of the free CO molecule.

\begin{figure}
\scalebox{0.77}{\includegraphics{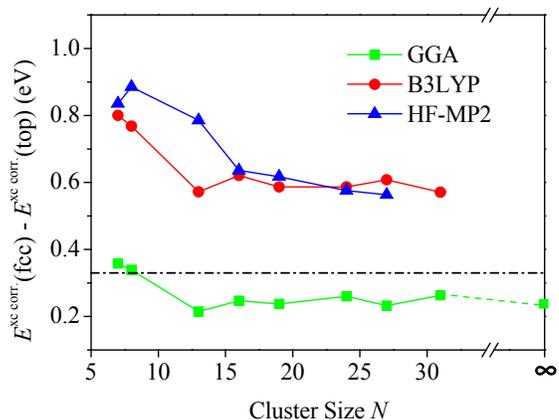}}
\caption{
The energy correction {\em difference} $E^{\rm xc\,corr.}({\rm fcc}) -
E^{\rm xc\,corr.}({\rm top})$. The dash-dotted line at 0.33 eV marks the
minimum xc correction required to obtain the correct top adsorption
site. The $N = \infty$ result for the GGA-LDA correction was obtained by
supercell calculations (LAPW/APW+lo).}
\label{fig-diff}
\end{figure}

Obviously, $E^{\rm xc\,corr.}$ 
is very different for different xc functionals. Interestingly, if
we evaluate \textit{differences}, e.g. for different adsorption sites,
these differences rapidly approach a constant value. Figure
\ref{fig-diff} shows the \textit{difference} of $E^{\rm xc\,corr.}$ for
CO at Cu(111) in the fcc and in the top adsorption sites.
Thus, xc-approximate total energy \textit{differences} of the extended
surface can be corrected through higher-level calculations for finite
clusters, and the cluster size where the xc energy correction term
$\Delta E^{\rm xc\,corr.}$ becomes constant determines the efficiency of
the local correction approach. For the GGA we also have the supercell
result at 1/9 monolayer coverage, where the adsorbate-adsorbate
interaction is negligible, and we added this $N=\infty$ result as well.
Figure \ref{fig-diff} demonstrates that $\Delta E^{\rm xc\,corr.}$ is
converged already for very small clusters (a 16 Cu atom cluster appears
to be sufficient). We emphasize that this applies to the
\textit{differences} and the LDA \textit{error}. For adsorption or
reaction energies such clusters are by far too small.
Apparently the LDA {\em error} is even shorter ranged than Kohn's 
nearsightedness concept \cite{Kohn-PRL} which
refers to interaction energies. 

Using the converged values of $\Delta E^{\rm xc\,corr.}$ enables us to
correct the LDA energy of the slab calculations. The GGA correction
decreases the wrong LDA preference for the fcc site, but it can not yet
change the energetic order of the hollow and top adsorption sites.
However, at the B3LYP level the top site is now more stable by
$0.21\pm0.03$\,eV. Interestingly, an almost identical value of
$0.28\pm0.04$\,eV is obtained at the HF-MP2 level. This confirms the
interpretation of earlier B3LYP studies
\cite{Gil-2003,Doll-2004,Doll-2006} that the main reason for the wrong
site preference of LDA and GGA functionals is the self-interaction error
(also present in the free CO molecule).

\begin{table}
\caption{
Adsorption energies (in eV) for low-coverage CO adsorption into
different high-symmetry adsorption sites at Cu(111) and Ag(111). The
values at the GGA and B3LYP levels are obtained through the xc energy
correction scheme, using the LDA numbers as reference. The energy of the
lowest energy structure is taken as energy zero.}
\begin{ruledtabular}
\begin{tabular}{l|l|l|l|l|l}
System  &xc      & top        & fcc       & hcp       & bridge\\ \hline
Cu(111) &GGA     & +0.13      & +0.01     & 0         & +0.05 \\
        &B3LYP   & 0          & +0.21     & +0.26     & +0.20 \\ \hline
Ag(111) &GGA     & 0          & +0.22     & +0.15     & +0.14 \\
        &B3LYP   & 0          & +0.42     & +0.43     & +0.35 \\
\end{tabular}
\end{ruledtabular}
\label{table-Cu-Ag}
\end{table}

Table \ref{table-Cu-Ag} gives the energies for all high-symmetry
adsorption sites at the Cu(111) and Ag(111) surfaces, namely top,
bridge, fcc, and hcp. For CO/Cu(111) the top site is now the most stable
adsorption site, and the optimum
diffusion energy barrier for the top-bridge-top pathway is 
0.20\,eV. For the other system, CO at Ag(111), already GGA yields the
correct top site as the most stable one \cite{Gajdos-2004,Mason-2004},
and it is interesting to verify that a higher-level xc treatment does
not spoil this description. As shown in Table \ref{table-Cu-Ag}, the
energetic order is indeed not changed at the B3LYP level, and the top
site is in fact further stabilized by 0.20\,eV, so that the diffusion
barrier for the top-bridge-top pathway is more than doubled: from
0.14\,eV (GGA) to 0.35\,eV (B3LYP).

\begin{figure}
\scalebox{0.77}{\includegraphics{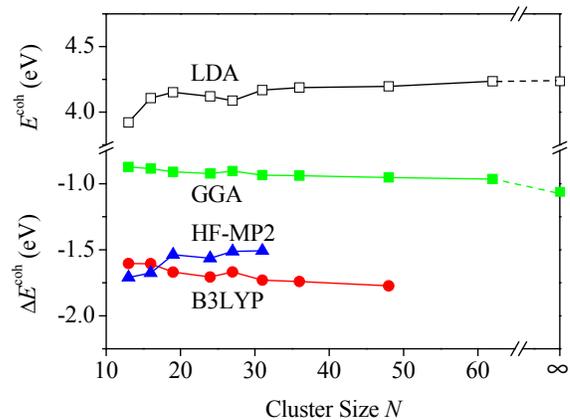}}
\caption{
Cohesive energy $E^{\rm coh.}$ of Cu (cf. eq. (\ref{bond-cut1})) for the LDA (black line, open squares), and $\Delta E^{\rm coh.}$
corrections with respect to the LDA for GGA, B3LYP, and HF-MP2. 
The $N = \infty$ result for the LDA and the GGA-LDA correction 
was obtained by crystal bulk calculations (LAPW/APW+lo).}
\label{fig-cohesiveE}
\end{figure}

The approach is not just applicable to localized perturbations, like an
adsorbate or a defect. It can also be used to study the bulk cohesive
energy. In this case, however, it is not possible to express the
correction in terms of energy differences so that cluster size and edge
effects cancel. We therefore now 
write the total energy of the many-atom system as sum over contributions
assigned to the individual atoms, $E = \sum_I E_I$. Here $E_I$ is the
energy contribution due to atom $I$. In the simplest, yet physically
meaningful approach for metals, $E_I$ is roughly proportional to $\sqrt{c}$, where
$c$ is the local coordination, i.e. the number of nearest neighbors
(see, e.g., \cite{Spanjaard-1990,scheffler-stampfl-2000}). For our
close-packed Cu clusters we then get
\begin{equation}
\label{bond-cut1}
E \;=\; N*E^{\rm atom} + (E^{\rm coh.}/\sqrt{12})
\sum_{c=1}^{12}\sqrt{c}* N_c \quad,
\end{equation}
where $E^{\rm atom}$ is the calculated total energy of the free atom at
the given xc level, $N$ the number of atoms in the cluster, $N_c$ the
number of atoms in the cluster that are $c$-fold coordinated, and
$E^{\rm coh.}$ corresponds to the cohesive energy expected for an
infinite size cluster. Figure \ref{fig-cohesiveE} shows how $E^{\rm
coh.}$ changes with cluster size and how it depends on the xc functional.

It can be seen that the correction $\Delta E^{\rm coh.}({\rm GGA - LDA})
= E^{\rm coh.}({\rm GGA})-E^{\rm coh.}({\rm LDA})$, as well as those for
the other xc functionals, is converged already for $N = 24$ clusters
with an uncertainty of $\pm 0.1$ eV. Hartree-Fock shows the same
convergence behavior as B3LYP but at a value of $-3.8$ eV. For 
the GGA we also performed calculations for the infinite crystal,
and this gives the
difference between the GGA and LDA cohesive energies as $-1.06$ eV, in
very close agreement to the converged value we get with our xc
correction scheme. Neef and Doll \cite{Doll-2006} had obtained 
values of $-1.05$ eV and $-2.04$ eV for the (GGA$-$LDA) and 
the (B3LYP$-$LDA) correction, respectively, which is rather
close to our results given in Fig. \ref{fig-cohesiveE}. However, 
this (B3LYP$-$LDA) correction is too large to match
experiment. The experimental cohesive energy for Cu is 3.49\,eV
\cite{janaf}, which differs by $-0.83$\,eV from the LDA cohesive energy.
Apparently, B3LYP is very bad in its description of Cu bulk. 
Also our cohesive energy at the HF-MP2 level (2.82 $\pm0.1$\,eV) is 
significantly smaller than the experimental value, but at this point we 
cannot rule out that the HF-MP2 convergence with cluster size 
is different to that of other treatments. In fact, whether or not
HF-MP2 should work for metals needs a deeper theoretical analysis.
A more detailed discussion will be given elsewhere. \cite{Fuchs-2007}

All results in Fig. \ref{fig-cohesiveE} were obtained for the LDA
lattice constant. Of course, we could have easily optimized the lattice
constants for the different treatments (or we could have shown the
equations of state for the different xc levels). However, this would
have complicated the graphs due to the intermingling of different
effects. 
The focus of the present work is on metal surfaces and bulk. However, the 
methodology proposed in this paper for correcting DFT-slab adsorption, cohesive, 
and diffusion energies is also applicable to semiconductors (e.g. H$_2$ at Si(001)
\cite{filippi02}), or ionic materials (e.g. NaCl). Not surprisingly, here the 
efficiency of the $\Delta$xc approach is even better (for details see 
\cite{Fuchs-2007}). The approach had been applied by Tuma and Sauer 
to protonation reactions in zeolites (combining DFT and HF-MP2).\cite{tuma}
This work is interesting as here the main source of error at the DFT level 
appears to be the lack of the long-range van der Waals tails, and by the 
correction scheme the authors could evaluate these van der Waals contributions 
to adsorption. 

In summary, we presented a scheme to locally correct the total energy
(or total energy differences) for errors contained in present-day local
or semi-local DFT functionals, e.g. the self-interaction and the lack of
van der Waals interactions. When looking at appropriate energy
differences, a smooth and rapid convergence of the xc correction with
cluster size is observed. This enables us to reach convergence at very
small cluster sizes or extra-polate to $N = \infty$. At these small
cluster sizes, that are treatable, e.g. with HF-MP2, it is only the xc
\textit{correction}, not the total energy, that is converged. The
approach is demonstrated by computing the energetic order of the
high-symmetry sites for the low-coverage adsorption of CO at Cu(111) and
Ag(111). The corrections to potential energy surfaces obtained with
LDA or GGA are found to be significant: For CO diffusion at Ag(111)
energy barriers are changed by more than a factor of two, and for CO
at Cu(111) even the topology is altered. Furthermore, the approach 
was applied to evaluate the cohesive energy of Cu bulk, enabling
us to perform HF-MP2 calculations (via a systematic extrapolation, starting from the LDA) for an extended system. In this paper we only addressed changes in energy, but for forces the correction is straightforward, analogous to eq. (\ref{xc-corr.}).

Discussions with Thorsten Kluener and Martin Fuchs are gratefully
acknowledged. Part of the work was supported by the Deutsche
Forschungsgemeinschaft (SFB658) and the Alexander von Humboldt foundation.


\begin{thebibliography}{99}
\bibitem{basis}
The LAPW/APW+lo basis set parameters are: $R_{\rm{MT}}^{\rm{Cu}}=
2.2$\,bohr,
$R_{\rm{MT}}^{\rm{Ag}}= 2.4$\,bohr, $R_{\rm{MT}}^{\rm{O}}= 1.0$\,bohr,
$R_{\rm{MT}}^{\rm{C}}= 1.0$\,bohr, $E^{\rm max}_{\rm wf} = 16$\,Ry,
$E^{\rm max}_{\rm pot} = 400$\,Ry. A $(4 \times 4 \times 1)$
Monkhorst-Pack grid was used for reciprocal space sampling of the $(3
\times 3)$ surface unit-cells.
\end{thebibliography}
\end{document}